\documentclass[prl,letterpaper,twocolumn,showpacs]{revtex4}

\usepackage{dcolumn}
\usepackage{amsmath,amsfonts}
\usepackage{graphicx}
\usepackage{color}
\usepackage{latexsym}
\usepackage{amssymb}
\usepackage{setspace}

\newcommand\comment[1]{} 

\begin{document}

\title{Enhancing Ionic Conductivity of Bulk Single Crystal
Yttria-Stabilized Zirconia by Tailoring Dopant Distribution}

\author{Eunseok Lee}
\author{Friedrich B. Prinz}
\author{Wei Cai}
\affiliation{Department of Mechanical Engineering, Stanford University,
Stanford, CA 94305}

\begin{abstract}
We present an \textit{ab-initio} based kinetic Monte Carlo model
for ionic conductivity in single crystal yttria-stabilized
zirconia.
Ionic interactions are taken into account by combining density
functional theory calculations and the cluster expansion method
and are found to be essential in reproducing the effective
activation energy observed in experiments.
The model predicts that the effective energy barrier can be
reduced by 0.15-0.25eV by arranging the dopant ions into a
super-lattice.
\end{abstract}


\date{\today}

\maketitle

%
Yttria-stablized zirconia (YSZ) is a widely used electrolyte in
solid oxide fuel cells (SOFC) and oxygen sensors because of its
high ionic conductivity at high temperatures~\cite{Cha}.
Driven by the need to reduce the operating temperature of SOFC,
much of the current research effort focuses on the design of new
solid electrolyte materials with significantly enhanced ionic
conductivity at intermediate temperatures~\cite{Hui,Steele}.
The presence of free surfaces in nanoscale thin films and
interfaces in hetero-epitaxial structures has been found to
enhance ionic
conductivity~\cite{Knoner,Souza,Sata,Kosaki,GBarriocanal,Schichtel}.
However, the effect of dopant distribution on the ionic
conductivity of bulk single-phase electrolytes has largely
remained unexplored, despite the fundamental importance of
dopant-vacancy interaction in ionic
transport~\cite{Stapper,Pietrucci} and the possibility of
tailoring dopant distributions by novel deposition
techniques~\cite{Chao}.

%
Atomistic simulations have the promise to become a useful design
tool for new electrolyte materials, by predicting the ionic
conductivity of candidate structures and elucidating the
fundamental transport
mechanisms~\cite{Sayle,Pennycook,Lee,Krishna,Rojana,Pietrucci}.
Unfortunately they are still limited in their length and time
scales.
For example, to accurately describe the long-range ionic
interactions in YSZ requires density functional theory (DFT)
models with relatively large supercells.
The high computational cost limits the time scale of \textit{ab
initio} molecular dynamics simulations to
picoseconds~\cite{Pennycook}.
Hence, a major challenge at present is to construct a kinetic
Monte Carlo (kMC) model that not only can access the macroscopic
time scale~\cite{Krishna,Rojana}, but also retains the accuracy of
DFT models in describing the ionic interactions.
In the pioneering kMC model for YSZ~\cite{Krishna}, ionic
interactions are ignored in the metastable states, i.e., all
possible states are sampled with uniform probability. While it
successfully predicts a maximum in the conductivity as a function
of doping concentration, the predicted temperature dependence is
significantly weaker than experiments, signaling the importance of
ionic interactions.
The lack of ionic interaction also makes this model unsuitable to
predict the effect of dopant distribution on ionic conductivity.

%
In this letter, we develop a kMC model for oxygen vacancy
diffusion in YSZ that faithfully captures the ionic interactions.
DFT calculations with supercell sizes significantly larger than
previous studies~\cite{Krishna,Predith} are performed to
accommodate long-range interactions, and the data are used to
construct a cluster expansion (CE) model.
%
%
KMC simulations using this model predicts an effective activation
energy that agrees better with experiments than the
non-interacting model.
The kMC simulations further predict that the maximum conductivity
is achieved when the Yttrium dopant ions are distributed as
$[100]$ lines and form a 2D rectangular super-lattice in the two
other directions.
%
%
The effective energy barrier in this structure is lower than the
random distribution by 0.15-0.25eV.

%
Ionic conduction in YSZ occurs through oxygen anion diffusion by
the vacancy mechanism.  The ionic conductivity is the averaged
effect of many vacancy jumps and can be predicted from a kMC
simulation over a sufficiently long time.
The degrees of freedom in our kMC model are the positions of the
oxygen vacancies, which hop on the simple cubic anion sublattice
of YSZ.
At each kMC step, the probability rates of all vacancy jumps to
their nearest neighbor positions are calculated by
$j=\nu_0\exp\left(-\frac{E_b}{k_BT}\right)$, where $k_B$ is
Boltzmann's constant, $T$ is temperature, and $E_b$ is the
activation energy barrier for each jump.  $\nu_0$ is a trial
frequency and is set to $10^{13}$~Hz~\cite{Krishna}.
At each step, only one event is selected based on the probability
rates of all possible events~\cite{KMC}.
From a long kMC simulation, the diffusion coefficient of the
vacancies center-of-mass is computed from the mean-square-displacement.
The ionic conductivity is then computed from the Nernst-Einstein
relation~\cite{Supple}.

\begin{figure}
\center
\includegraphics[width=3in]{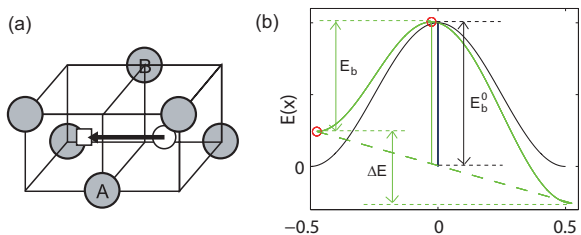}
\caption{(a) In the non-interacting model $E_b$ depends only on
whether cations $A$ and $B$ are Zr or Y. $E_b = 0.58$, $1.29$ or
$1.86$~eV if within pair $A$-$B$ there are 0, 1, or 2 Y
cations~\cite{Krishna}. Filled circles represent cations, open
circle represents oxygen anion, and open square represents oxygen
vacancy. (b) In our interacting model, the energy difference
$\Delta E$ between the initial and final state is assumed to
modify the energy profile of vacancy jump by superimposing a
linear function.
%
%
} \label{fig:Ebmodel}
\end{figure}
%

A fundamental input to the kMC simulation is the energy barriers
for vacancy jumps, $E_b$, which depends on the ionic
configurations around the jumping vacancy.
In a previous model~\cite{Krishna}, $E_b$ is assumed to depend
only on the chemical species of the two cations closest to the
jumping vacancy, as shown in Fig.~\ref{fig:Ebmodel}(a).
Because the energy barrier of every jump equals that of the
reverse jump, one can show that all metastable states in this
model must have identical energy.
Hence we will refer to it as the {\it non-interacting} model.
Experimental and computational data have suggested that
interactions play an important role in ionic
conduction~\cite{Stapper,Pietrucci}.
To account for interactions, we use the kinetically resolved
activation (KRA) model~\cite{Ven1,Ven2}, in which $E_b =
f(E_b^0,\Delta E)$, where $E_b^0$ is the ``kinetically resolved''
barrier when the two metastable states happen to have identical
energy, and $\Delta E$ is the energy difference between the two
states.
We take the energy barriers in the non-interacting
model~\cite{Krishna} as our $E_b^0$.
The function $f$ is often approximated by $E_b = E_b^0 - \Delta
E/2$~\cite{Ven1,Ven2}.
%
%
Here we use a slightly better approximation for function $f$ by
assuming that the energy landscape between the two metastable
states has a sinusoidal shape when $\Delta E = 0$, and is modified
by a linear term when $\Delta E\neq 0$~\cite{Supple}.
%
%
Hence our task of specifying the energy barrier $E_b$ is reduced
to an accurate description of the energy difference $\Delta E$.

Because the ionic interactions in YSZ are long ranged, they can
only be captured accurately by DFT calculations in relatively
large supercells. It is impossible to perform DFT calculations for
all ionic configurations sampled by the kMC simulation.
Instead, we use the cluster expansion method (CEM) to limit the
necessary number of DFT calculations.
In CEM, every metastable state in the kMC simulation can be
uniquely mapped to a spin configuration $\{s_i\}$ of an Ising
model \cite{Predith}.
%
%
%
The energy as a function of ionic configuration can be expressed
by a cluster expansion~\cite{Tepesch}
\begin{equation}
 E(\{s_i\})=\sum_{\alpha}V_{\alpha}\phi_{\alpha}(\{s_i\})
\label{eq:lineareq}
\end{equation}
where $V_{\alpha}$ is called effective cluster interaction (ECI)
for cluster $\alpha$, and $\phi_{\alpha}= \prod_{i\in \alpha} s_i$
is the cluster function involving all spin variables belonging to
cluster $\alpha$.
%
%
Eq.~(\ref{eq:lineareq}) is used to compute the energy of any
metastable state encountered by the kMC simulation, prior to which
the ECI coefficients $V_\alpha$'s are fitted to a DFT data set.


To construct the {\it ab initio} database for the fitting, we
performed DFT calculations for 140 randomly chosen ionic
configurations within a $3[100]\times3[010]\times3[001]$ YSZ
supercell, using the Vienna Ab-initio Simulation Package
(VASP)~\cite{VASP}.
This supercell contains 108 cation sites (Zr or Y) and 216 anion
sites (O or V$_{\rm O}$) and is significantly larger than previous
studies~\cite{Krishna,Predith}, in order to accurately account for
long range Coulombic and elastic interactions.
%
%
K-point sampling is limited to the $\Gamma$-point considering the
large size of supercell.
The volume and the shape of the supercell are allowed to relax
together with the ionic positions.
A high energy cut-off of 520~eV is used to avoid Pulay stress.
Each ionic configuration takes $\sim 10^4$ CPU-hours to be fully
relaxed.


%
The number of clusters needs to be significantly truncated for
robust fitting, otherwise the cluster expansion model can be
overly-adapted to the fitted data set~\cite{Rosen}.
First, we only keep the clusters in which any two spins are
separated by less than 1.5$a_0$, where $a_0$ is the lattice
parameter of YSZ.
Second, we only keep clusters that involve up to 3 spins.
Accounting for the translational and rotational symmetries, 192
independent clusters survive this truncation.
%
%
%
%
%
%
For further truncation, a Monte Carlo algorithm is used to select
$n_c$ clusters out of 192 possible clusters by minimizing the
cross validation score~\cite{Predith}.
To measure the predictive power of CEM, we separate the DFT data
into two sets.  Set I contains 100 data points and are used to fit
the ECI's.  Set II contains 40 data points are used to benchmark
the CEM's predictions.
The root mean square difference between the DFT energies and the
CEM's predictions per cation in Set I and Set II are defined as
the error of fitting and the error of prediction, respectively.

%
\begin{figure}
\center
\includegraphics[width=2.6in]{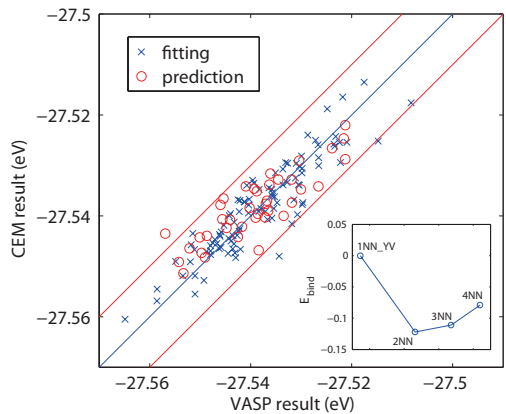}
\caption{(color online) The 140 data points are divided into two
sets: set I contains 100 data points ($\times$) and is used to fit
CEM, set II contains 40 data points ($\circ$) and is not included
in the fit.  Energies predicted from CEM are plotted against
energies computed from DFT using VASP, in units of eV per cation.
The inset shows the effective binding energy between Y and oxygen
vacancy predicted by CEM.}
\label{fig:CEMresult}
\end{figure}

These two errors have different dependence on $n_c$.
For example, when $n_c = 97$, the error of fitting is $0.0006$~eV
while the error of prediction is $0.012$~eV. The large difference
between the two errors means that the fitted CEM has entirely lost
its predictive power if $n_c$ is too large.
Only when $n_c \leq 9$, both errors are the same, and decrease
with increasing $n_c$.
Hence, in this work, the optimal choice is $n_c =
9$~\cite{Supple}, where both the error of fitting and the error of
prediction is $0.005$~eV, as shown in Fig.~\ref{fig:CEMresult}.
This error is small enough for our kMC simulations and is smaller
than a previous study~\cite{Predith}.
%
%
To our knowledge this is the first time the predictive power of an
interaction model for YSZ has been demonstrated by monitoring the
error of prediction.
%
%
Fig.~\ref{fig:CEMresult}(b) shows the effective binding energy
between an Y ion and an oxygen vacancy predicted by the CEM model.
The preference of the oxygen vacancy to the second nearest
neighbor site of Y is clearly seen, consistent with previous
experimental and theoretical works~\cite{Krishna,Rojana}.

\begin{figure}
\center
\includegraphics[width=3.2in]{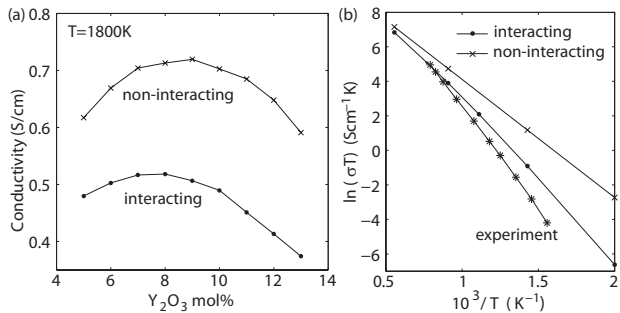}
%
\caption{
(a) Conductivity of YSZ single crystal predicted by kMC
simulations at $T = 1800$~K as a function of doping concentration.
(b) Predicted temperature dependence of YSZ conductivity at
8-mol\% doping concentration.
Y dopant cations are assumed to be randomly distributed.
Experimental data is reproduced from \cite{Ikeda}.
} \label{fig:Conductivity}
\end{figure}

The fitted CEM allows us to compute the energy difference $\Delta
E$ between the two states before and after a vacancy jump, which
modifies the energy barrier $E_b$ of the jump.
Using this energy barrier model, we performed kMC simulations in
$3[100]\times3[010]\times3[001]$ YSZ supercells in which the
doping concentration varies from $5\%$ to $13\%$.
The ionic conductivity at each doping concentration is computed by
averaging over 40 randomly generated Y distributions.
As shown in Fig.~\ref{fig:Conductivity}(a), the ionic conductivity
(at 1800~K) is maximum at 8-mol\%, consistent with earlier
experimental~\cite{Ioffe,Ikeda} and theoretical
results~\cite{Krishna,Rojana}.
Fig.~\ref{fig:Conductivity}(b) is the Arrhenius plot of ionic
conductivity at 8mol\% doping concentration. The predicted
activation energy is 0.74eV at high $T$ and 0.85eV at low $T$, in
much better agreement with experiments (0.85-1.0~eV) ~\cite{Ikeda}
than the non-interacting model (0.59~eV)~\cite{Krishna}. The
remaining difference with experiments may be due to the error in
$E_b^0$ taken from~\cite{Krishna}.
The neglect of activation entropy in vacancy jumps does not affect
the conclusion because it does not change the slope of the
Arrhenius plot.

%



The conductivity results shown in Fig.~\ref{fig:Conductivity} are
the averaged value over 40 random Y configurations, the standard
deviation over which is about $19\%$ of the average.
This indicates that the ionic conductivity is sensitive to the
spatial distribution of Y cations and poses the question: what is
the optimal Y distribution that maximizes ionic conductivity?
To answer this question, we have performed kMC simulations for a
variety of Y distributions, in which the Y cations are segregated
into either spherical clusters, (001) layers, or [100] rods.
The simulation results suggest two design principles that
ultimately guide us to the optimal Y distribution.

When all Y cations in the supercell are segregated into a
spherical cluster, the ionic conductivity is actually lower than
the random distribution (by $27\%$ at $1800$~K), contrary to
the prediction based on the non-interacting model~\cite{Lee}.
This is because the interaction between Y cations and oxygen
vacancies, as shown in Fig.~\ref{fig:CEMresult}, attracts the
vacancies next to the cluster.
To diffuse over long distances and contribute to the ionic
conductivity, vacancies must detach from the Y clusters. This
requires overcoming a binding energy of $\sim0.12$~eV, which reduces
the ionic conductivity.

Here the reduction of ionic conductivity is caused by the increase
of spatial variation of the potential energy.  Based on this
result, we can formulate {\em design principle} I: the optimal Y
distribution should minimize the energy variation of the
metastable states as oxygen vacancies jump along the conduction
direction.
In this work, we focus on conduction along the [100] (i.e.\ $x$)
direction.  A promising candidate structure is to have Y cations
segregated into planes, so that each (001) cation layer is either
completely filled by Y, or completely filled by Zr.
Due to translational invariance, the potential energy from
cation-vacancy interaction remains constant after an oxygen
vacancy jumps in the [100] direction~\footnote{The vacancy-vacancy
interaction can still induce a small energy variation.}.
The layered structure can be fabricated using thin film deposition
techniques such as PLD~\cite{Chao}.

Unfortunately, the layered structure also has a lower conductivity
than the random distribution (by $59\%$ at $1800$~K).
This is surprising because one might expect enhanced conductivity,
as the non-interacting model would predict, due to the existence
of Y-free channels.
The reduction in conductivity is caused by the segregation of
oxygen vacancies to the two anionic layers immediately next to the
Y layer. Because there is always a first nearest neighbor (1nn) Y,
vacancy diffusion in this layer experiences a high energy barrier
(with $E_b^0 = 1.29$~eV).
Given that vacancies prefer to be the second nearest neighbors
(2nn) of Y, as shown in Fig.~\ref{fig:CEMresult}, it is somewhat
surprising that the vacancies segregate to the nearest anionic
layer.
This problem is resolved by noticing that when the vacancy becomes
the 1nn of two Y cations, it becomes the 2nn to four Y cations on
the same cation layer.
This result motivates our {\em design principle} II: the optimal Y
distribution should not induce a high oxygen vacancy density in
the first nearest neighbor sites of any Y cations.
%

It turns out that the design principles I and II can be
simultaneously satisfied when Y cations are segregated into [100]
lines.  When Y lines are well separated from each other, kMC
simulations show that the oxygen vacancy density is peaked at 2nn
sites surrounding the Y lines.  To enhance conductivity, we would
like to pack more Y lines per unit volume in order to enhance the
overall vacancy density.  But when the distance between Y lines
becomes too small, oxygen vacancies start to occupy 1nn sites
around Y, deteriorating the ionic conductivity.

\begin{figure}
\center
\includegraphics[width=3.2in]{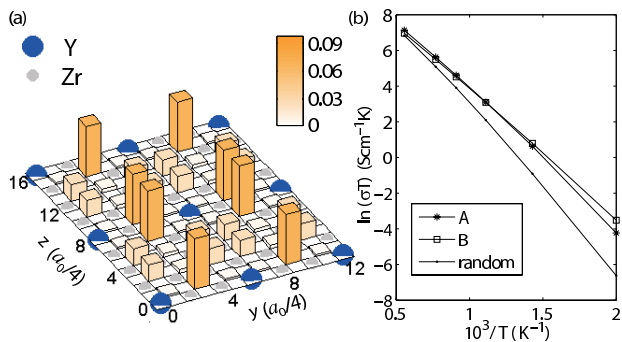}
\caption{(color) (a) Vacancy distribution in structure B at $T =
500$~K - blue and gray circles indicate Y and Zr, respectively.
Darker color in the orange columns corresponds to higher vacancy
concentration. (b) Temperature dependence of the conductivity at
different Y distributions. Random Y distribution corresponds to a
doping concentration of 8mol\%.
%
%
} \label{fig:Dist_Eb_eff}
\end{figure}

We examined a variety of 2D structures formed by [100]
Y-segregated lines.
The optimal structure at $1800$~K is a square lattice with
periodicity 1.5$a_0$ in $y$ and $z$ directions (structure A).
Interestingly, the optimal structure at $500$~K is different; it
is a rectangular lattice with periodicity 1.5$a_0$ and 2$a_0$ in
$y$ and $z$ directions, respectively (structure B).
In both structures, the vacancy concentration is high in fast
diffusing channels ($E_b^0 = 0.58$~eV) away from 1nn sites of Y,
see Fig.~\ref{fig:Dist_Eb_eff}(a).

Fig.~\ref{fig:Dist_Eb_eff}(b) plots the temperature dependence of
conductivity for structures A and B.  Their activation energy is
around $\sim0.6$~eV, significantly lower than the random
distribution.
Compared with the random Y distribution (at 8mol\%), the ionic
conductivity of structure A is enhanced by a factor of 1.35 at
1800~K, 11 at 500~K and 86 at 300~K. For structure B, the
enhancement factor is 22 at 500~K and 532 at 300~K.
This result provides a theoretical upper limit of ionic
conductivity that can be achieved by rearranging dopants in YSZ.
%
%

%
%

In summary, we have developed an \textit{ab-initio} based kMC
model of the vacancy diffusion in bulk YSZ that accurately
accounts for the ionic interactions.
The predicted ionic conductivity shows much better agreement with
experiments regarding the temperature dependence than the
non-interacting model.
The model predicts strong dependence of ionic conductivity on the
spatial distribution of dopant cations.
The maximum conductivity is reached when Y cations are arranged
into a rectangular superlattice of [100] lines.
Fabrication of this structure is challenging, but may be feasible
with novel deposition techniques.
The method presented here can be easily applied to other solid
electrolytes, in which the optimal dopant microstructure may be
different from that in YSZ and may be easier to synthesize.
%
%

This work is partly supported by the DOE/SciDAC project on Quantum
Simulations of Materials and Nanostructures.  E. Lee acknowledges
support from the Samsung Scholarship Foundation. F. B. Prinz
acknowledges support from the DOE-EFRC: CNEEC (Award No.
DE-SC0001060)

\end{document}